\documentclass[reprint,twocolumn,showpacs,amsmath,amssymb,superscriptaddress,aps,prl]{revtex4-1}
\usepackage{amsmath}
\usepackage{amsfonts}
\usepackage{amssymb}
\usepackage{graphicx}
\usepackage{epsfig}
\usepackage{color}
\usepackage{empheq}
\usepackage{float}
\usepackage{epstopdf}
\usepackage{hyperref}
\usepackage{lineno}

\newcommand{\RR}{\right}
\newcommand{\LL}{\left}
\newcommand{\m}{\mathrm}
\newcommand{\dg}{\dagger}
\newcommand{\eref}[1]{Eq.~(\ref{#1})}
\newcommand{\fref}[1]{Fig.~\ref{#1}}

\newcommand{\vari}{}
\newcommand{\varib}{}

\begin{document}

\title{Entangled massive mechanical oscillators}

\author{C. F. Ockeloen-Korppi}
\affiliation{Department of Applied Physics, Aalto University, P.O. Box 15100, FI-00076 AALTO, Finland}
\author{E. Damsk\"agg}
\affiliation{Department of Applied Physics, Aalto University, P.O. Box 15100, FI-00076 AALTO, Finland}
\author{J.-M. Pirkkalainen}
\affiliation{Department of Applied Physics, Aalto University, P.O. Box 15100, FI-00076 AALTO, Finland}
\author{A. A. Clerk}
\affiliation{Institute for Molecular Engineering, University of Chicago, Chicago, IL, 60631, USA}
\author{F. Massel}
\affiliation{Department of Physics and Nanoscience Center, University of Jyv\"askyl\"a, P.O. Box 35 (YFL), FI-40014 University of Jyv\"askyl\"a, Finland}
\author{M. J. Woolley}
\affiliation{School of Engineering and Information Technology, UNSW Canberra, ACT, 2600, Australia}
\author{M.~A.~Sillanp\"a\"a}
\email[]{mika.sillanpaa@aalto.fi}
\affiliation{Department of Applied Physics, Aalto University, P.O. Box 15100, FI-00076 AALTO, Finland}

\begin{abstract}
\textbf{An entangled quantum state of two or more particles or objects exhibits some of  the most peculiar features of quantum mechanics. Entangled systems cannot be described independently of each other even though they may have an arbitrarily large spatial separation. Reconciling this property with the inherent uncertainty in quantum states is at the heart of some of the most famous debates in the development of quantum theory \cite{EPR}. Nonetheless, entanglement nowadays has a solid theoretical and experimental foundation, and it is the crucial resource behind many emerging quantum technologies. Entanglement has been demonstrated for microscopic systems, such as with photons \cite{Aspect,Heidmann87TwoMSq,Kimble92TwoMSq,Bowen2002}, ions \cite{Wineland09Ent}, and electron spins \cite{Hanson2015Bell}, and more recently in  microwave and electromechanical devices \cite{Martinis06Ent,Schoelkopf2010Ent,LehnertEnta2013}. For macroscopic objects \cite{Leggett1980,Polzik2001Ent,Diamonds2011,Awschalom2015}, however, entanglement becomes exceedingly fragile towards environmental disturbances. A major outstanding goal has been to create and verify the entanglement between the motional  states of slowly-moving massive objects. Here, we carry out such an experimental demonstration, with the moving bodies realized as two micromechanical oscillators coupled to a microwave-frequency electromagnetic cavity that is used to create and stabilise the entanglement of the  centre-of-mass motion of the oscillators \cite{ClerkEnt2014,WoolleyBAE,2BAE}. We infer the existence of entanglement in the steady state by combining  measurement of correlated mechanical fluctuations with an analysis of the microwaves emitted from the cavity. Our work qualitatively extends the range of entangled physical systems, with implications in quantum information processing, precision measurement, and tests of the limits of quantum mechanics.}
\end{abstract}

\maketitle

\varib{There exist several proposals for how cavity optomechanical setups could be used to entangle the motional quantum states of two massive mechanical oscillators, see e.g., Refs.~\cite{OptoEntang,Heidmann2005,Meystre2013Sq,Clerk13WangC,ClerkEnt2014,Vitali15Ent}. In such setups, two movable mirrors are incorporated into a resonant optical cavity, and radiation pressure forces inside the cavity can be tailored such that the motion of the mirrors becomes highly correlated and even entangled. This approach does not require any direct interaction between the moving masses, allowing them to be spatially separated. A correlated reduction of noise below the thermal level in the motion of two mechanical oscillators have previously been demonstrated in experiment \cite{YamaguchiSqu2014,Marin2016TwoSqu}, but not near the quantum level as required for entanglement.}

\begin{figure}[h]
\includegraphics[width=8.5cm]{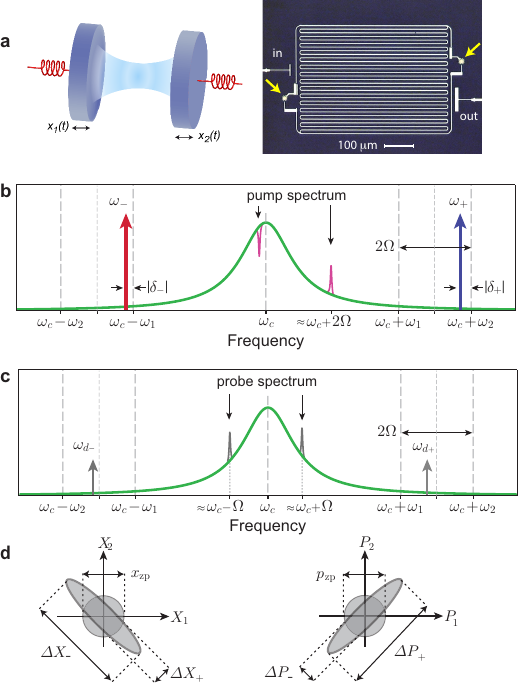}
\caption{\emph{Creating and detecting motional entanglement.} \textbf{a}, Schematic of two vibrating mirrors coupled via an electromagnetic cavity (left), and a micrograph showing the microwave optomechanical device consisting of a superconducting transmission line resonator, whose opposite ends are connected by two mechanical drum-type oscillators marked with arrows (right).  The device is fabricated on a quartz chip out of aluminium. \textbf{b}, Spectral picture of the pump microwave frequencies (see text) applied about the cavity spectrum. \textbf{c}, Two additional weak probe tones are applied in order to reconstruct the $X_+$ collective mechanical quadrature using a two-mode back-action evading measurement. \textbf{d}, Illustration of the correlations in two-mode squeezing in terms of fluctuations (shaded) of the quadrature amplitudes. Left, the sum of $X$ quadratures of the two oscillators fluctuates less than the zero-point level (circle). Right, the difference of $P$ quadratures is similarly localized.}
\label{Fig1}
\end{figure}

Our work is based on a series of proposals \cite{Meystre2013Sq,Clerk13WangC,ClerkEnt2014} for using reservoir engineering  to stabilise two cavity-coupled mechanical oscillators into a steady state that is entangled. The recipes are extensions of an approach  used in recent work to squeeze the motion of a single oscillator  \cite{SchwabSqueeze,Squeeze,TeufelSqueeze}.  An oscillator with  frequency $\omega_1$ and position operator $x_1(t) = X_1(t) \cos \omega_1 t + P_1(t) \sin \omega_1 t$, is squeezed if the variance of either quadrature amplitude operators $X_1$ or $P_1$ is smaller than the quantum zero-point fluctuation level. Introducing a second oscillator with frequency $\omega_2$ and position operator $X_2$ expressed in terms of quadratures as above, one can introduce four collective quadrature operators, \varib{$X_\pm = \frac{1}{\sqrt{2}} \LL( X_2 \pm X_1 \RR)$ and $P_\pm = \frac{1}{\sqrt{2}} \LL( P_2 \pm P_1 \RR)$.} The state corresponding to either the variances of $X_+$ and $P_-$, or $X_-$ and $P_+$, being reduced below the quantum zero-point fluctuations level is a canonical entangled state known as the two-mode squeezed state.
\varib{Such a state leads to violations of local realism, and is analogous to the state considered in} the EPR paradox \cite{EPR}, see \fref{Fig1}d. Specifically, the state is entangled if $\langle X_+^2 \rangle+ \langle P_-^2 \rangle < 1$, a criterion commonly referred to as the Duan inequality \cite{Duan}.

\varib{In our experiment, a single driven cavity mode is used both to prepare a correlated state of two mechanical oscillators as well as to directly measure fluctuations in the $X_+$ collective quadrature (via a two-mode BAE measurement) \cite{ClerkEnt2014,WoolleyBAE,2BAE}; we find $\langle X_+^2 \rangle \simeq 0.41 \pm 0.04$. While a direct measurement of $P_{-}$ is not possible, by constraining system parameters and analyzing the full output spectrum of our cavity, we are also able to infer its fluctuations, $\langle P_-^2 \rangle \simeq 0.42 \pm 0.08$ (see e.g.~Ref.~\cite{SchwabSqueeze}).} We hence have an entangled state with $\langle X_+^2 \rangle+ \langle P_-^2 \rangle \simeq 0.83 \pm 0.13 < 1$ of two slowly moving massive oscillators each consisting of approximately $10^{12}$ atoms.


As shown in \fref{Fig1}a, we use a microwave-frequency realization of cavity optomechanics involving two micromechanical drum oscillators \cite{Teufel2011b}, and a superconducting on-chip circuit acting as the electromagnetic cavity (frequency $\omega_\m{c}$). The oscillators' positions affect the total  capacitance, and hence modulate the frequency of the cavity. This creates an effective radiation pressure interaction as with optical cavities and mirrors. In order to generate two-mode squeezing and entanglement, we pump the system with two strong pump microwave tones at the frequencies $\omega_-$ and $\omega_+$, below and above (respectively) the cavity frequency as shown in \fref{Fig1}b.

To describe this system, we initially assume for simplicity that the two oscillators have equal single-photon radiation-pressure coupling strengths $g_0$, and that the pumps are applied at the red and blue sideband frequencies $\omega_- = \omega_c - \omega_1$ and $\omega_+ = \omega_\m{c} + \omega_2$, respectively. Details of the derivations, including non-idealities, are discussed in the supplement \cite{supplement}. The strong pumps enhance the radiation-pressure interaction, \varib{yielding many-photon} coupling rates $G_\pm = g_0 \alpha_\pm$, where $\alpha_\pm$ are the field amplitudes induced in the resonator due to the pump tones at $\omega_{\pm}$.  We also introduce the mechanical Bogoliubov modes which are obtained by a two-mode squeezing transformation on the original mechanical annihilation operators, viz.~ $\beta_1 = b_1 \cosh r+b^\dag_2 \sinh r$, and $\beta_2 = b_2 \cosh r+b^\dag_1 \sinh r$, where $\tanh r = G_+/G_-$. \varib{Defining $\Omega = (\omega_2-\omega_1)/2$, and working in a rotating frame (at $\omega_{\m{c}} + \Omega$ for the cavity, $ (\omega_2+\omega_1)/2$ for each mechanical oscillator), the linearized optomechanical Hamiltonian is:}
\begin{equation}
\label{eq:Hamilt}
\begin{split}
H = &- \Omega a^\dg a + \Omega \LL( \beta_2^\dg \beta_2 - \beta_1^\dg \beta_1 \RR)  \\
& + \mathcal{G} \LL[ a^\dg \LL( \beta_1 +  \beta_2 \RR) + a \LL( \beta_1^\dg +  \beta_2^\dg \RR)\RR] \,.
 \end{split}
 \end{equation}
\varib{Here $\mathcal{G} = \sqrt{G_-^2 - G_+^2}$ is an effective optomechanical coupling rate.} This Hamiltonian is essentially that in Ref.~\cite{ClerkEnt2014}, but with the pump tones set as in Ref.~\cite{Clerk13WangC}.
It describes cooling of the Bogoliubov modes by cavity cooling towards their ground state, which corresponds to a stabilised, two-mode squeezed state of the bipartite mechanical system. In contrast to dynamical protocols, e.g.~Refs.~\cite{Diamonds2011,LehnertEnta2013}, the system hence stays entangled indefinitely. Here, non-degenerate mechanical frequencies are essential, so that both Bogoliubov modes are efficiently cooled by different frequency components of the cavity. 


Although the pumping frequencies correspond to those in Ref.~\cite{Clerk13WangC}, we stress that in our scheme both pumps necessarily couple to both oscillators as in \cite{ClerkEnt2014}, entailing that the rotating-wave approximation involves only a modest requirement \varib{that the cavity linewidth} $\kappa \ll \omega_{1,2}$, as is well satisfied in the experiment. \varib{Note that we will work with $\Omega \lesssim \kappa$, so that both $\beta_1$ and $\beta_2$ have appreciable coupling to the cavity via \eref{eq:Hamilt}.} If there are asymmetries in the single-photon couplings $g_1$Ê and $g_2$ for the two oscillators, or if the pump tones are detuned (by the amounts $\delta_\pm$, see \fref{Fig1}b, \varib{where we also define} $\Delta = (\delta_+ - \delta_-)/2$) from the mechanical sidebands, there are additional terms in the Hamiltonian \eref{eq:Hamilt}, and one might expect that the steady-state entanglement is reduced (see \cite{supplement} for the full model). However, we find numerically that one can greatly compensate for asymmetries in couplings by optimising the pump detunings $\delta_\pm \neq 0$. We find, as well, that with the current set of parameters, it is beneficial to red-pump the lower-frequency oscillator 1. We will follow these practices in the experiment.

As displayed in \fref{Fig1}c, an essential part of the entanglement verification strategy consists of two-mode back-action evading (BAE) detection \cite{WoolleyBAE,2BAE} operated in the same cavity mode, \varib{allowing for mapping the mechanical motion to the output field}. This involves two relatively weak probe tones applied at \varib{$\omega_{\m{d}\pm} \approx \omega_\m{c} \pm (\omega_2 + \omega_1)/2$}, approximately in the middle of the sideband frequencies. In order to preserve the same rotating frame for creation and detection of the two-mode squeezing, we strictly require $\omega_{\m{d}+} -\omega_{\m{d}-} = \omega_{+} -\omega_{-}$, ideally up to complete phase coherence between the tones. In a manner similar to the pumps, the probes induce effective couplings $g_\pm = g_0 \alpha^\m{d}_\pm$, with the amplitudes $\alpha^\m{d}_\pm$, which in the ideal two-mode back-action evading case are equal. Since we are using the same cavity mode for both creating the entanglement via the pumps and detecting it, the pump spectra and probe spectra need to be independent. \vari{This is achieved by ensuring that the mechanical contributions to the output cavity spectrum from the pump and probe tones have negligible spectral overlap. Hence, the faithful reconstruction of the $X_+$ collective quadrature spectrum from the probe signal is possible.} 
In contrast to BAE detection of single-mode squeezing \cite{SchwabSqueeze}, both the pumps and probes can be set to optimal frequencies for the creation and detection of two-mode squeezing (see \fref{Fig1}b,c).

In our device the two oscillators are far separated by 600 microns, they have no direct coupling, and the system is well described by \eref{eq:Hamilt}. We use the fundamental drum modes of the oscillators with the resonance frequencies $\omega_1/2\pi \simeq 10.0$ MHz and $\omega_2/ 2\pi \simeq 11.3$ MHz, and linewidths $\gamma_1/2\pi \simeq   106 $~Hz and $\gamma_2/ 2\pi  \simeq 144$ Hz, respectively. The microwave cavity, with the frequency $\omega_\m{c}/2\pi \simeq  5.5$~GHz, has separate input and output ports. All the input signals are applied through a  port coupled weakly at the rate $\kappa_\m{Ei}/2\pi \simeq  60$~kHz, whereas the output is strongly coupled at $\kappa_\m{Eo}/2\pi  \simeq 1.13$~MHz. The cavity also has internal losses at the rate $\kappa_\m{I}/2\pi \simeq 190$~kHz, and all the loss channels sum to the total linewidth $\kappa/2\pi \simeq  1.38$~MHz. 
We  find that our fabrication process can produce basically identical single-photon couplings, $g_1/g_2 \simeq 0.98$ for two oscillators of different frequencies, in fact, this is more than sufficient for the purpose of generating entanglement \varib{since numerically we find that an asymmetry up to $\sim 20$ \% can be compensated via detunings.}

The  motion of the mechanical oscillators is measured via the  power scattered from the applied microwave tones, both pumps and probes, \varib{ by their interaction with the oscillators}. We collect this weak signal using standard techniques, including a low-noise cryogenic microwave amplifier followed by room-temperature signal analysis. A sequence of calibrations, described in detail in the supplement \cite{supplement} is important for the experiment. First, based on a standard thermal calibration using a single red-detuned tone, the mechanical modes are found to thermalize down to the equilibrium phonon occupation  numbers $n_1^{\m{T}} \simeq 41$ and $n_2^{\m{T}} \simeq 30$ for oscillators 1 and 2, respectively, at the base temperature $\simeq 14$ mK of the dry dilution refrigerator. These values imply the initial variances of the collective quadratures $\langle X_\pm^2\rangle^\m{T}$ and $\langle P_\pm^2\rangle^\m{T} \simeq 36$.

%
\begin{figure}
\centering
\includegraphics[trim = 0mm 10mm 3mm 10mm, clip,width=8.5cm]{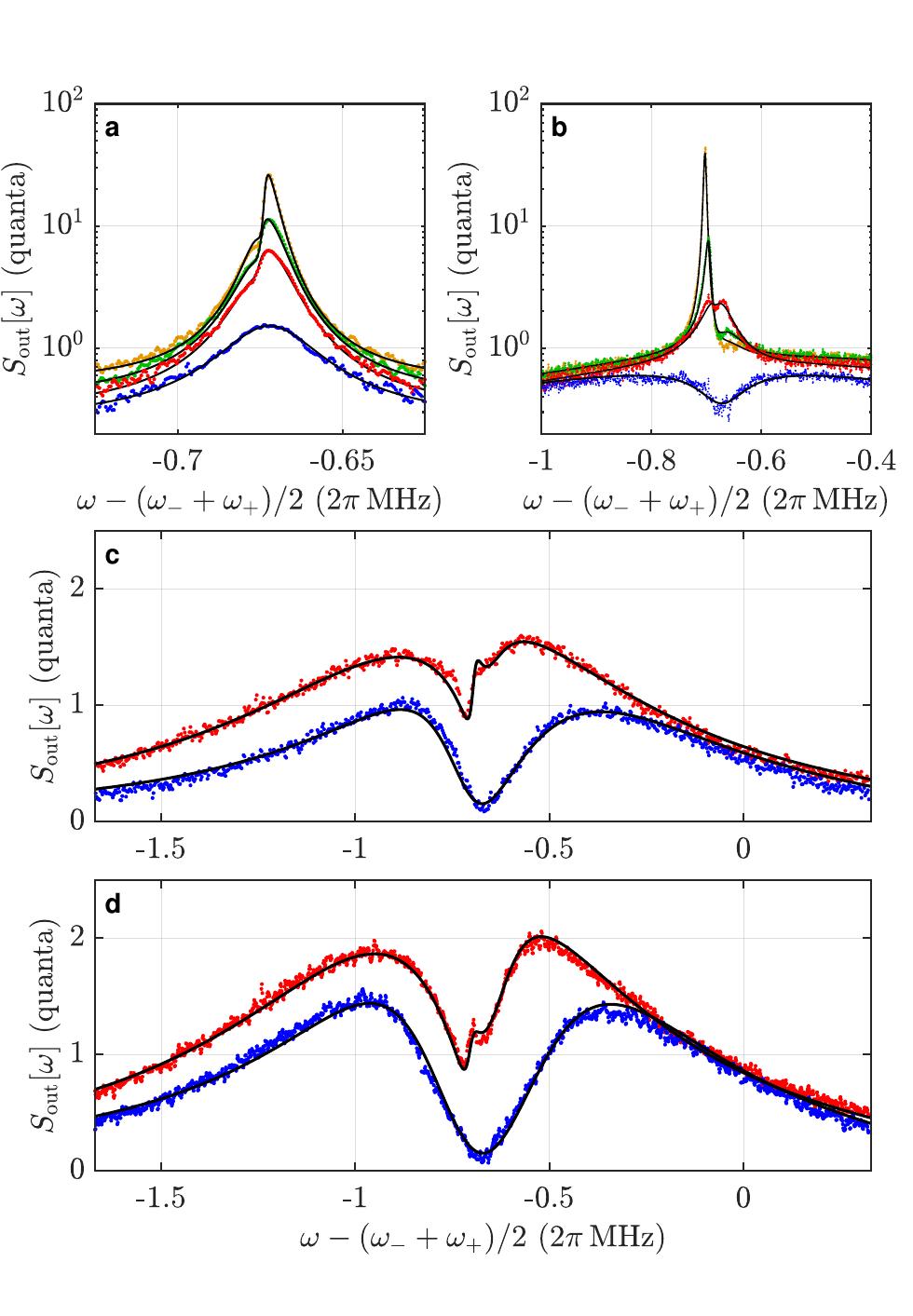}
\caption{\emph{Pump spectra under two-tone driving.} \textbf{a}, The pump amplitudes are as $G_-/2\pi \simeq 103$ kHz, and varying $G_+/2\pi \simeq 82, \, 74, \, 67$ kHz from top to bottom. \textbf{b}, $G_-/2\pi \simeq 201$ kHz, $G_+/2\pi \simeq 135$ kHz, and varying detuning $\Delta /2\pi \simeq [19.4, \, 9.6,  \, 0.1]$ kHz from left to right. \textbf{c}, Labelled dataset \emph{C} in the following, with $G_-/2\pi \simeq 278$ kHz, $G_+/2\pi \simeq 166$ kHz. \textbf{d}, Dataset \emph{D}, having higher pump powers $G_-/2\pi \simeq 332$ kHz, $G_+/2\pi \simeq 210$ kHz. The blue lines, shown for reference, are the sideband cooling calibrations run for oscillator 1. Theory curves are given by the solid lines. The remaining parameters are listed in the supplementary \cite{supplement}.}
\label{FigPump}
\end{figure}

We proceed with standard sideband cooling of each mechanical oscillator separately using a single red-detuned pump. This allows characterization of the behaviour of the system under intense pumping. Importantly, it calibrates the gain of the detection system for the later interpretation of the spectrum under two-tone pumping, as well as  the effective coupling of the red-detuned tone. 
The goal of calibrating the probe tones is to use the total power in the probe spectra as a straightforward thermometer for the quadratures. Similar to the single red tone case, we run a thermal calibration with both probe tones on that allows us to determine the collective occupation number measured at a small probe power. Second, a power sweep calibration of the probes connects a given signal strength to the quadrature \varib{variance}.

Next we discuss the main experiment that uses two pairs of tones, namely the pumps and the probes. The pump tones are used to create entanglement, and the probe tones enable tomography of the collective mechanical state. First, we focus on the spectrum due to the pump tones. In \fref{FigPump}, we display the pump output spectra $S_{\m{out}}[\omega]$ from the cavity under several pumping conditions, given in absolute units determined via the  gain calibration. This spectrum is one piece of information available for characterizing the entanglement of the two oscillators, since it allows for the inference of the effective temperatures of the three reservoirs. Our theoretical modeling uses standard input-output theory for electromagnetic cavities, treating the pump and probe tones as effectively \vari{belonging to independent modes}. The variances of the collective quadratures, with a given set of parameters, can be evaluated within the same framework. 

  
%
\begin{figure}
\centering
\includegraphics[trim = 8mm 25mm 3mm 0mm, clip,width=8.5cm]{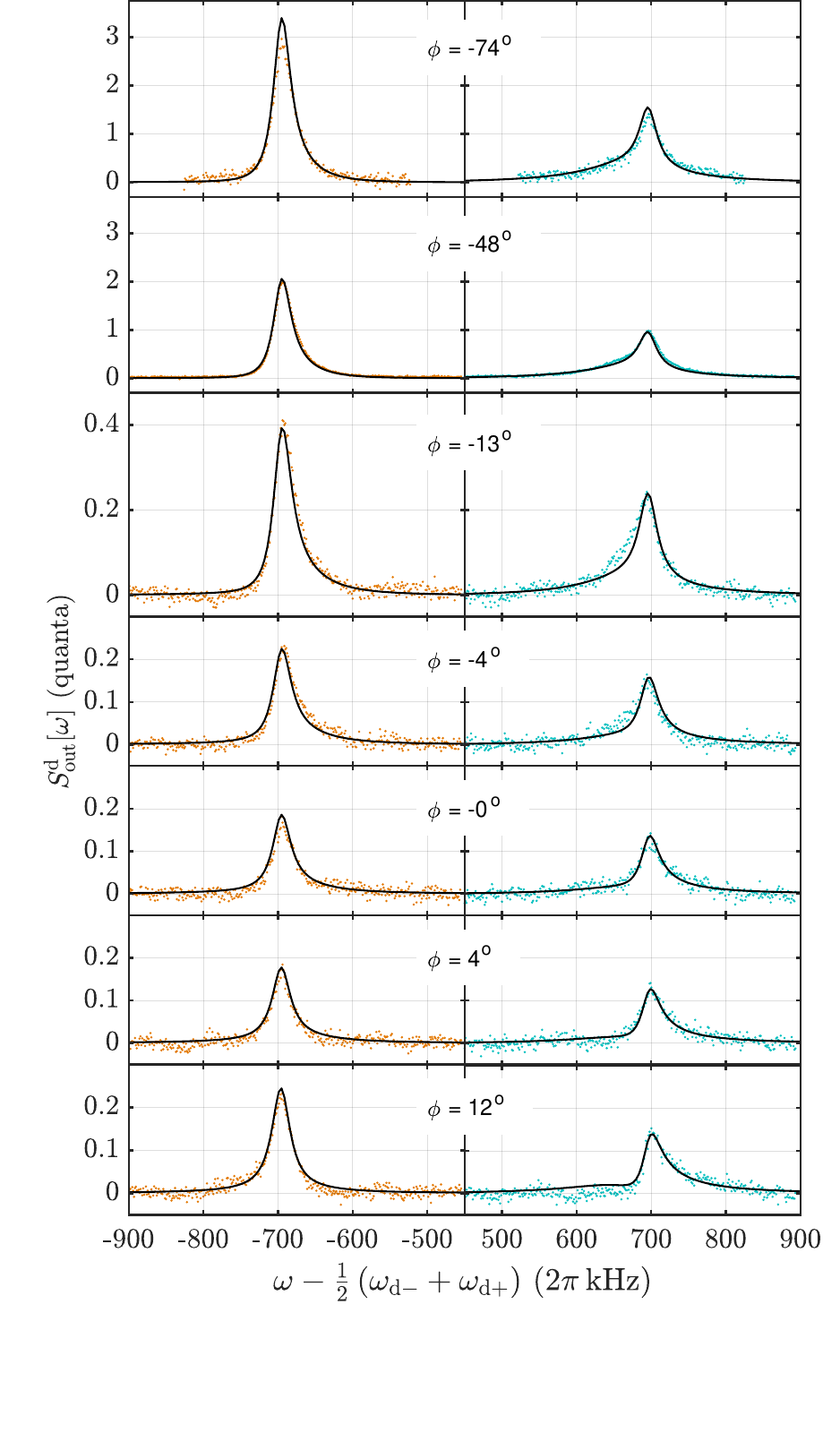}
\caption{\emph{Two-mode back-action evading readout.} The probe output spectrum recorded with the same parameters as the pump spectrum in \fref{FigPump}c (dataset \emph{C}), i.e., $G_-/2\pi \simeq 278$ kHz, $G_+/2\pi \simeq 166$ kHz. The probe phases are written in the panels, and the solid lines are theoretical predictions. The four lowermost panels exhibit two-mode squeezing below \vari{the quantum zero-point fluctuation level}.
The two uppermost panels, having a different vertical scale, present hot quadratures. }
\label{Fig2BAE}
\end{figure}

Now consider the spectrum resulting from the probe tones. The ability to control the relative phase $\phi$ of the two probe tones allows us to infer the variance of a general collective quadrature $X_+^\phi = X_+ \cos\phi + P_+  \sin\phi$. 
Written in terms of the spectra $S_{\m{X}_+}[\omega]$ and $S_{\m{P}_+}[\omega]$ of $X_+$ and $P_+$, respectively, the measured spectrum is then proportional to $S_{\m{X}_+^\phi}[\omega] =S_{\m{X}_+}[\omega]  \cos^2 \phi + S_{\m{P}_+}[\omega]  \sin^2 \phi$. 
In the measurement, the probe signal is visible as peaks on top of the pump spectrum at the frequencies $\sim \omega_\m{c} \pm \Omega$ on either side of the cavity (see \fref{Fig1}c). In \fref{Fig2BAE} we display the measured probe spectra $S^\m{d}_{\m{out}}[\omega]$, having subtracted the background measured in the absence of the probe tones, corresponding to dataset \emph{C}. The probe power at $g_\pm/2\pi \simeq 40$ kHz was kept much smaller than the pump power.
The theoretical model, shown with the same parameters for all curves, is in excellent agreement with the experiment, including the positions and markedly non-Lorentzian lineshapes of the peaks, all of which depend on the phase. The unusual shapes physically \vari{arise due to the fact} that the two oscillators are pumped in a very unequal manner, and they exhibit individual optical springs which add up to the collective spectrum. 
We infer the quadrature \vari{variance} $\langle (X^\phi_+)^2 \rangle $ from the total integrated area of the peaks. \vari{This method, as opposed to relying on the particular shapes of the peaks,}
is insensitive to the phase drift of the microwave sources occurring during the data acquisition.
Within a typical integration time for one curve of approximately 30 minutes, $\phi$ can drift several degrees, leading to slight departures from the theoretically determined curves.

\begin{figure*}
\centering
\includegraphics[trim = 0mm 10mm 10mm 0mm, clip,width=13cm]{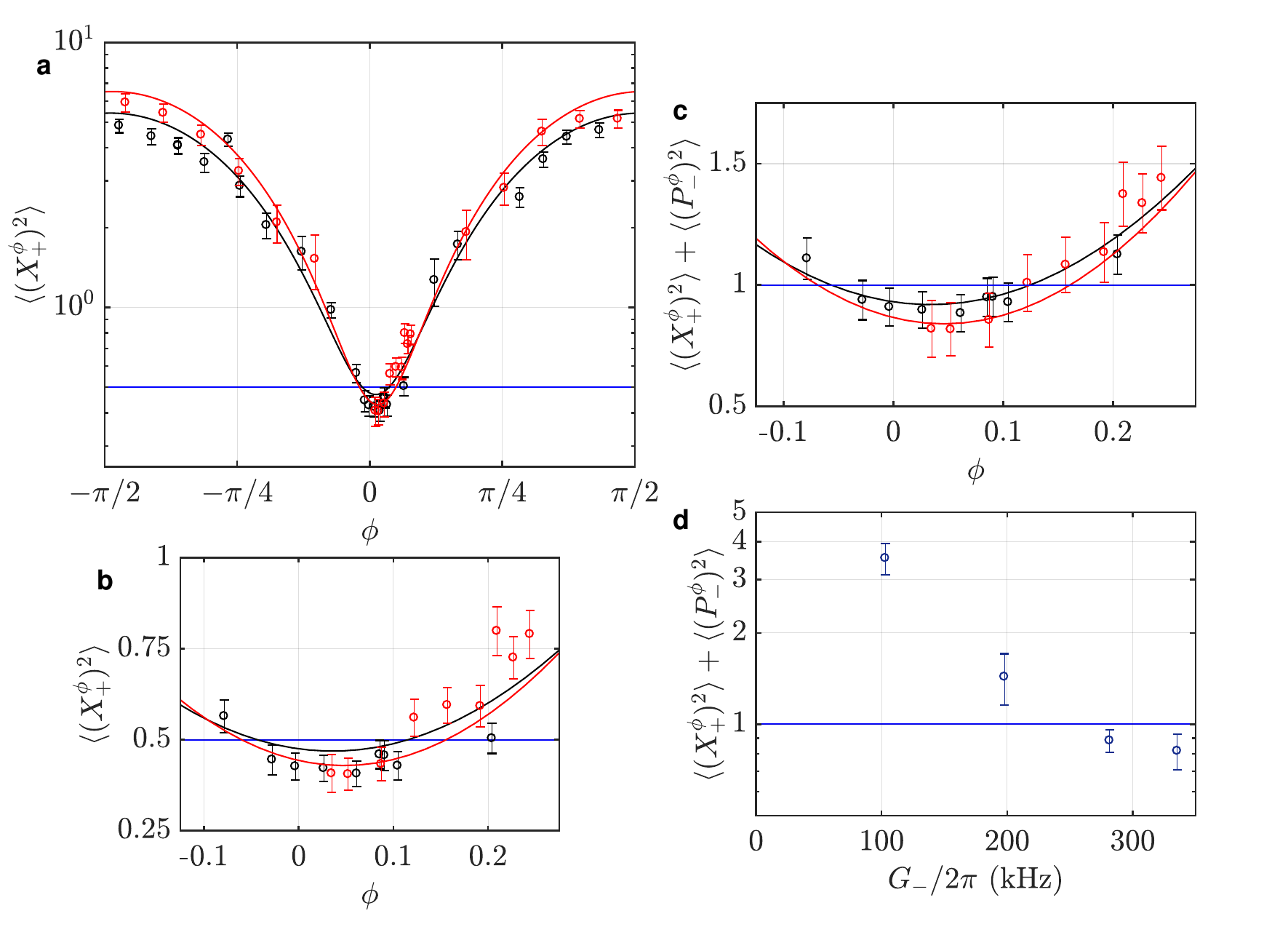}
\caption{\emph{Fluctuations of collective quadratures.}  \textbf{a}, The $X_+^\phi$ quadrature variance measured by the probe spectra. The black circles come from the dataset \emph{C}, and red circles correspond to dataset \emph{D}. \textbf{b}, Zoom-in of \textbf{a}. \textbf{c}, The Duan quantity for entanglement as a function of the probe tone phase. \textbf{d}, The Duan quantity, for the optimal value of $\phi$, as a function of the strength of the red-detuned pump tone. The solid lines coloured similar to the datasets are theoretical fits using the bath temperatures determined by the pump spectra. The blue horizontal line marks the quantum zero-point fluctuations level.}
\label{FigEntangle}
\end{figure*}

From our modeling \cite{supplement}, we confirm that the probe peak area faithfully reproduces  the quadrature \vari{variances} following our calibration as described above. This rigorously holds if the probe powers are perfectly matched, i.e., \varib{$\alpha^{\rm d}_+ = \alpha^{\rm d}_-$}, which was calibrated without the pump tones. However,  for the theory fits in \fref{Fig2BAE}, we need to \vari{determine an imbalance} of $g_-/g_+ \simeq 1.055$, which we attribute to a shift of the cavity frequency when the strong pumps are present, causing a reduction of the blue probe tone in the system involving \varib{an input filter with a steep slope at the blue side}. The reduced $g_+$, given that $g_-$ stays constant, means that the probe area will under-estimate the quadrature \vari{variance}, here by $\simeq 18$ \%, a number obtained numerically from the model \cite{supplement}. We will hence scale up the quadrature variances inferred from the probe areas by this percentage. A similarly low blue pump power as compared to the best estimate is observed in the pump spectra as well, but the red pump matches the calibration, supporting the handing of the imbalance as described.  We  emphasize that the probe inference does not assume anything about the mechanical oscillators or the dynamics induced by the pumps, but only assumes an understanding of the dynamics associated with the probe tones.


In  \fref{FigEntangle}a,b we display a \varib{tomographic measurement} of the $X_+^\phi$ collective quadrature, with 95 \% statistical confidence intervals. In the optimal case of $\phi \simeq 4^\m{o}$ minimizing the variance in \fref{Fig2BAE}, we obtain  $\langle (X^\phi_+)^2 \rangle \simeq 0.41 \pm 0.04$; that is 0.9 dB below vacuum. Several points fall well below the quantum zero-point noise level in both datasets \emph{C} and \emph{D}. Since the best theoretical fit to the measured probe spectra is obtained with dataset \emph{C}, we base our main claims on this data. We believe that dataset \emph{D} was subject to larger phase drifts during the data acquisition. Generally, the spectra and responses are highly sensitive to the parameters, and it is remarkable that we  \vari{reach such good agreement} between theory and experiment with a fixed set of parameters describing different measurements (pump, probe, and linear response \cite{supplement}).


Now consider the measurement of the variance of the $P_-$ quadrature, needed for examining the Duan  criterion and verifying quantum entanglement. The two-mode BAE probe detection, as mentioned, does not couple to $P_-$ or $X_-$. We therefore use the other source of information available, namely the pump spectrum, and then combine this information with that provided by the probe detection. A least-squares fit to an analytical expression describing the pump spectrum, using the three bath temperatures as adjustable parameters, combined with the aforementioned calibrations, allows \vari{the evaluation of these variances}. The fits are shown in \fref{FigPump}, displaying \vari{an excellent} agreement to the experiment. 
For dataset \emph{C}, we obtain the variance  $\langle P_-^2 \rangle \simeq 0.45  \pm 0.08$. For $X_+$ quadrature, we similarly get $\langle X_+^2 \rangle \simeq 0.46  \pm 0.08$, close to the value obtained from the direct BAE detection method described above (for the confidence intervals, see next section). Given our knowledge of system parameters and the dynamics of this scheme, the two quadratures are expected to have variances within 5 \% of one another \cite{supplement}, providing additional evidence for the value of $\langle P_-^2 \rangle$ based on BAE detection.


The error analysis of the probe measurement uses straightforward error propagation of the experimental calibrations, and of a statistical error from integrating the probe peak area. For the pump spectrum, the analysis is  complicated because it involves more parameters, \vari{some of which can sensitively affect the steady-state entanglement.}
Here we adopt an error analysis method known as the Bayesian Monte Carlo method, similar to Ref.~\cite{SchwabSqueeze}, to rigorously infer the parameters including uncertainties and correlations. The method generates a sample of the parameter distribution for which the theory model agrees with the measured pump spectra within the statistical uncertainty \cite{supplement}.
We sample the posterior distributions of all parameters, and use the distributions to estimate the  confidence limits of the $P_-$ quadrature variance. We obtain \cite{supplement} that at 96 \% probability, $\langle P_-^2 \rangle < 0.5$ in case of the data in \fref{Fig2BAE} (dataset \emph{C}). This approach also yields the most likely value $\langle P_-^2 \rangle \simeq 0.42 \pm 0.08$ that  agrees with the values obtained above, but is determined independently.


The best estimate of the Duan quantity is given by combining all the information, namely $\langle X_+^2 \rangle$ from probe detection, and that for $\langle P_-^2 \rangle$ as explained above, with equal-weight error bars. We therefore conclude that individually the fluctuations of $X_+$ and $P_-$ are below the quantum zero-point fluctuations level, and in particular that when summed, they satisfy the Duan bound for entanglement, $\langle X_+^2 \rangle + \langle P_-^2 \rangle < 1$ for dataset \emph{C} at \vari{98 \% probability, exceeding} the standard $2 \sigma$ confidence. The conclusion is further supported by dataset \emph{D} (red in \fref{FigEntangle})  \cite{supplement}, satisfying the Duan bound even more \vari{strongly at $> 99$ \% probability}. 

The entangled mechanical oscillators we have prepared can find practical use in sensitive measurements. Combined with two-mode back-action evading measurements, \varib{squeezed optical inputs, and parametric amplification}, they would allow for infinitely precise reconstruction of classical forces driving the oscillators. This has implications for gravitational wave detection, and metrology more broadly. Fundamentally, the entanglement of massive mechanical oscillators establishes a new regime for experimental quantum mechanics. The correlations we have created between the motion of two oscillators are so strong that they can be said to share a channel of spooky action, thereby measuring one would influence the other non-locally. In the future one could carry out quantum teleportation of motional states or test Bell inequalities with massive mechanical objects.

\bibliographystyle{naturemag}


\bibliography{/Users/masillan/Documents/latex/MIKABIB}

\textbf{Acknowledgements} We would like to thank Sorin Paraoanu and Ian Petersen for useful discussions. This work was supported by the Academy of Finland (contracts 250280, 308290, CoE LTQ, 275245) and by the European Research Council (615755-CAVITYQPD). We acknowledge funding from the European UnionÕs Horizon 2020 research and innovation program under grant agreement No. 732894 (FETPRO HOT). The work benefited from the facilities at the Micronova Nanofabrication Center and at the Low Temperature Laboratory infrastructure.


\end{document}